# Carrier-Resolved Photo Hall Measurement in World-Record-Quality Perovskite and Kesterite Solar Absorbers

Oki Gunawan[1*], Seong Ryul Pae[2], Douglas M. Bishop[1], Yun Seog Lee[1,3], Yudistira Virgus[1], Nam Joong Jeon[4], Jun Hong Noh[4,5], Xiaoyan Shao[1], Teodor Todorov[1], David B. Mitzi[6], Byungha Shin[2*]

[1]IBM T. J. Watson Research Center, P.O. Box 218, Yorktown Heights, NY 10598, USA.
[2]Dept. Materials Science and Engineering, Korea Advanced Institute of Science and Technology, 291 Daehak-Ro, Yuseong-Gu, Daejeon 34141, Republic of Korea.
[3]Dept. Mechanical and Aerospace Engineering, Seoul National University, 1 Gwanak-Ro, Gwanak-Gu, Seoul 08826, Republic of Korea
[4]Div. of Advanced Materials, Korea Research Institute of Chemical Technology, 141 Gajeong-ro, Yuseong-Gu, Daejeon 305-600, Republic of Korea.
[5]School of Civil, Environmental and Architectural Engineering, Korea University, Seoul 136-713, Republic of Korea.
[6]Dept. of Mechanical Engineering and Material Science, Duke University, Durham, NC 27708, USA.

*Corresponding authors: Oki Gunawan: ogunawa@us.ibm.com, Byungha Shin: byungha@kaist.ac.kr

**Majority and minority carrier properties such as type, density and mobility represent fundamental yet difficult to access parameters governing semiconductor device performance, most notably solar cells. Obtaining this information simultaneously under light illumination would unlock many critical parameters such as recombination lifetime, recombination coefficient, and diffusion length; while deeply interesting for optoelectronic devices, this goal has remained elusive. We demonstrate here a new carrier-resolved photo-Hall technique that rests on a new identity relating hole-electron mobility difference ($\Delta\mu$), Hall coefficient ($h$), and conductivity ($\sigma$): $\Delta\mu = \left(2 + d\ln h / d\ln \sigma\right) h\sigma$, and a rotating parallel dipole line ac-field Hall system with Fourier/lock-in detection for clean Hall signal measurement. We successfully apply this technique to recent world-record-quality perovskite and kesterite films and map the results against varying light intensities, demonstrating unprecedented simultaneous access to the above-mentioned parameters.**

Hall effect is one of the most important characterization techniques for electronic materials and has become the basis of fundamental advances in condensed matter physics, such as the integer and fractional quantum Hall effect (*1, 2*). The technique reveals fundamental information about the majority charge carrier, i.e. its type (*P* or *N*), density and mobility. In a solar cell, the majority carrier parameters determine the overall device architecture, width of the depletion region and bulk series resistance. The minority carrier properties, however, determine other key parameters that directly impact overall device performance, such as recombination lifetime ($\tau$), electron-hole recombination coefficient ($\gamma$) and minority carrier diffusion length ($L_D$). Unfortunately, the standard Hall measurement only yields majority carrier information. Extraction of minority carrier information often relies on multiple characterization methods, for instance, a combination of $L_D$ measurement from bias-dependent quantum efficiency (QE) and lifetime from time-resolved photoluminescence (TR-PL) (*3*). Separate measurement techniques typically utilize different illumination levels and may require different sample configurations, thus presenting additional complications in the analysis.

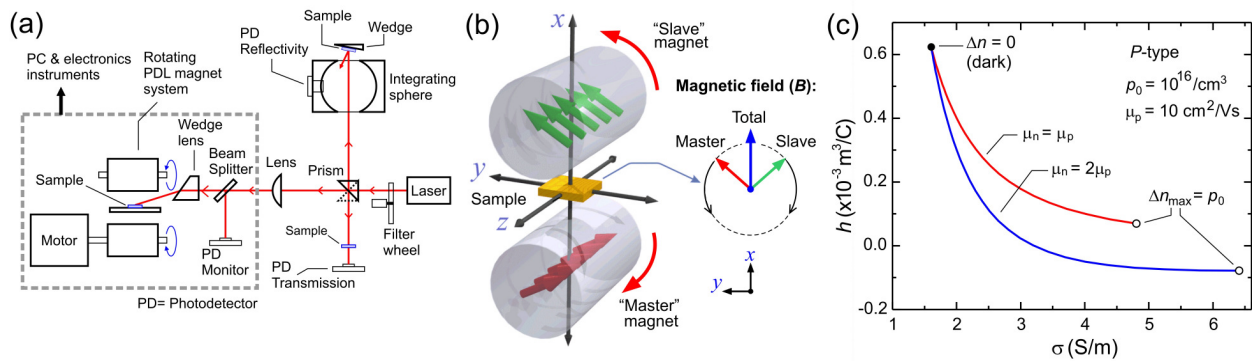

**Fig. 1 The carrier-resolved photo Hall measurement:** (a) The parallel dipole line (PDL) photo Hall setup for a complete photo-Hall experiment. (b) The rotating PDL magnet system that generates unidirectional and single harmonic ac magnetic field at the center (see animation in Movie S1). (c) Two systems with the same majority mobility ($\mu_p$) but different minority mobility ($\mu_n$) yield different Hall coefficient-conductivity ($h$-$\sigma$) curves.

Extraction of reliable minority carrier information is particularly sought after in the study of organic-inorganic hybrid perovskites, a material garnering intense attention due to fast progress in achieving high performance solar cells, with the current record power conversion efficiency (PCE) of 22.7% (*4*), and its applications as optoelectronics devices such as light emitting diodes (*5*) and photodetectors (*6*). Full understanding of charge transport properties of the perovskite will help unveil the operating principles of high efficiency solar cells, thereby guiding further improvement. A second thin-film semiconductor absorber of high interest—i.e., kesterite $Cu_2ZnSn(S,Se)_4$, (CZTSSe) which comprises only earth abundant metals—provides a more stable platform for photovoltaic (PV) and photoelectrochemical devices compared to current generation perovskites, and has also garnered intense attention (*7*). Attempts to push efficiency above 12.6% have stalled, however, and require detailed knowledge of majority/minority carriers within the absorber. Studies to collect both majority/minority carrier properties for high-performance films and crystals in these materials have been attempted, but require a wide range of experimental techniques (*3,8-16*) (cf. Table S3 and S4). A better approach in extracting the minority/majority carrier information simultaneously from a single sample is of high interest for all key photovoltaic technologies and other associated optoelectronic devices.

In this work we present a carrier-resolved photo-Hall measurement technique capable of simultaneously extracting both majority and minority carrier mobilities, densities and subsequent derivative parameters ($\tau$, $L_D$ and $\gamma$) as a function of light intensity. This technique rests on two key breakthroughs: (1) a new identity equation for the photo-Hall effect that yields the mobility difference, (2) a high sensitivity, rotating *parallel dipole line* (PDL) ac field Hall system (*17-19*) with its Fourier spectral decomposition and visual lock-in detection for clean Hall signal measurement. Besides the photo-Hall measurement, optical measurements to calculate the absorbed photon density ($G$) (e.g., transmission and reflectivity) can also be done in the same setup as shown in Fig. 1(a) (see also supplementary materials (SM) E), making this a self-contained system.

In the classic Hall measurement without illumination one can obtain three parameters for majority carriers—i.e., the (i) type ($P$ or $N$), from the sign of the Hall coefficient $h$; (ii) density: $p = 1/he$, and (iii) mobility: $\mu = h\sigma$, where $\sigma$ is the (longitudinal) conductivity and $e$ is the electron's charge. The key issue in the photo-Hall transport problem is to solve for three unknowns at a given illumination: hole and electron mobility $\mu_p$, $\mu_n$ and their photo-carrier densities $\Delta n$ and $\Delta p$, which are equal under steady state. Unfortunately, we only have two measured quantities: $\sigma$ and $h$. The key insight in solving this problem is illustrated in Fig. 1(c). Consider two systems with the same majority density ($p_0$) and mobility ($\mu_p$) but different minority mobility ($\mu_n$). When these systems are excited with the same photo-carrier density, $\Delta n_{max}$, they will produce different $h$-$\sigma$ curves – due to the increasing role of the minority carrier mobility to the total conductivity under illumination –even though they start from the same point in the dark. Therefore the characteristics of the $h$-$\sigma$ curves, specifically the slope ($dh/d\sigma$) at any given point, contain information about the systems' mobility. We show that the mobility difference $\Delta\mu = \mu_p - \mu_n$ at any point in the $h$-$\sigma$ curve is given as (see SM C):

$$\Delta\mu = \left(2 + \frac{d\ln h}{d\ln\sigma}\right)h\sigma .$$

(1)

Note that $h$ and $\sigma$ are experimentally obtained against varying light or absorbed photon density $G$ (or $\Delta n$), e.g. $h(G)$ and $\sigma(G)$, but fortuitously this $G$ or $\Delta n$ dependence cancels out of equation 1. We note that the term $d\ln h/d\ln\sigma$ has special experimental meaning as shown in the perovskite analysis example later. This equation applies to both carrier types and assumes that the mobilities are fairly constant around the $h$-$\sigma$ point of interest. Using the known two-carrier expressions: $\sigma = e(p\mu_p + n\mu_n)$ and $h = (p - \beta^2 n)/(p + \beta n)^2 e$, where $p$ and $n$ are hole and electron densities and $\beta = \mu_n/\mu_p$ is the mobility ratio, we can completely solve the photo-Hall transport problem ($\mu_p$, $\mu_n$ and $\Delta n$), e.g. for $P$-type material:

$$\beta = \frac{2\sigma(\Delta\mu - h\sigma) - \Delta\mu^2 e\,p_0 \pm \Delta\mu\sqrt{e\,p_0}\sqrt{\Delta\mu^2 e\,p_0 + 4\sigma(h\sigma - \Delta\mu)}}{2\sigma(\Delta\mu - h\sigma)}$$

(2)

$$\Delta n = \frac{\sigma(1 - \beta) - \Delta\mu\,e\,p_0}{\Delta\mu\,e\,(1 + \beta)} ,$$

(3)

Finally we obtain: $\mu_p = \Delta\mu/(1-\beta)$ and $\mu_n = \beta\mu_p$. Note that we need to know $p_0$ from a dark measurement. We refer to Eq. 1-3 as the "$\Delta\mu$" model. As shown later, when the $h$ data are noisy, the solutions could have large fluctuations. To address this case, one can use constant average majority and minority mobility that have been determined using the "$\Delta\mu$" model above and solve for $\Delta n$. We refer to this model as "$\mu_{PN}$" model whose formula are given in SM C.

The second key ingredient involves obtaining clean Hall measurements. Unfortunately, in many PV films, high sample resistance ($R > 10$ G$\Omega$) as in perovskites and low mobility ($\mu < 1$ cm$^2$/Vs) as in kesterites produce very noisy Hall signals. Therefore, ac Hall techniques coupled with Fourier analysis and lock-in detection are crucial. We recently developed a high sensitivity ac-field Hall system based on a rotating PDL magnet system (17-19). The PDL system is a recently-discovered magnetic trap that harbors a new type of field confinement effect that generates a magnetic camelback potential along its longitudinal axis (20). This effect is used to optimize the field uniformity (see SM B). The PDL Hall system consists of a pair of diametric cylindrical magnets separated by a gap. One of the magnets (the "master") is driven by a motor and the other (the "slave") follows in the opposite direction. This system produces a *unidirectional* and *single harmonic field* at the center where the sample resides (see Fig. 1(b) and Movie S1), which forms the basis for a successful photo-Hall experiment.

To demonstrate this technique, we performed the carrier-resolved photo-Hall experiment on a (FA,MA)Pb(I,Br)$_3$ perovskite film, fabricated using the same method that produced world-record solar cell efficiency (21) but with further process optimization. A companion device in the same batch yielded PCE of 21.2% (SM A). The device is a Hall bar with sample thickness $d$=0.55 μm (Fig. 2(b) inset). First we measured the sample in the dark and obtained the majority carrier properties: *P*-type, $p_0$=(3.1±0.7)×10$^{11}$/cm$^3$ and $\mu_p$=(17±4) cm$^2$/Vs. Then, we performed the measurement under several laser intensities (wavelength λ=525 nm, up to ~5 mW/cm$^2$). An example of longitudinal ($R_{XX}$) and transverse ($R_{XY}$) magnetoresistance (MR) traces at the highest light intensity are shown in Fig. 2(a). The $R_{XY}$ trace shows the expected Hall signal with a Fourier component at the same frequency as the magnetic field $B$ ($f_{REF}$) [Fig. 2(a)(vii)]. The desired Hall signal $R_H$ is obtained using numerical lock-in detection based on a reference sinusoidal signal with the same phase as $B$ [Fig. 2(a)(iv)] (17). The $h$ and σ values are then calculated from $R_H$ and $R_{XX}$ (see SM D). We also observe a second harmonic component at $2f_{REF}$ in the $R_{XY}$ Fourier spectrum [Fig. 2(a)(vii)], which is also evident in the original $R_{XY}$ trace as a double frequency oscillation [Fig. 2(a)(iii)]. This component is not the desired Hall signal and thus is rejected. It arises from another magnetoresistance effect (22), which is stronger in $R_{XX}$ [Fig. 2(a)(ii)]. It also appears in $R_{XY}$ because of $R_{XX}$-$R_{XY}$ mixing due to the finite size of the Hall bar contact arms. This *highlights the importance* of inspecting the Hall signal Fourier spectrum and using lock-in detection for obtaining clean Hall signals, as opposed to simple amplitude measurement. Furthermore, the lock-in detection also assures that the extracted Hall signal has the same phase (in-phase) with $B$, and thus rejecting the parasitic out-of-phase Faraday emf signal (17).

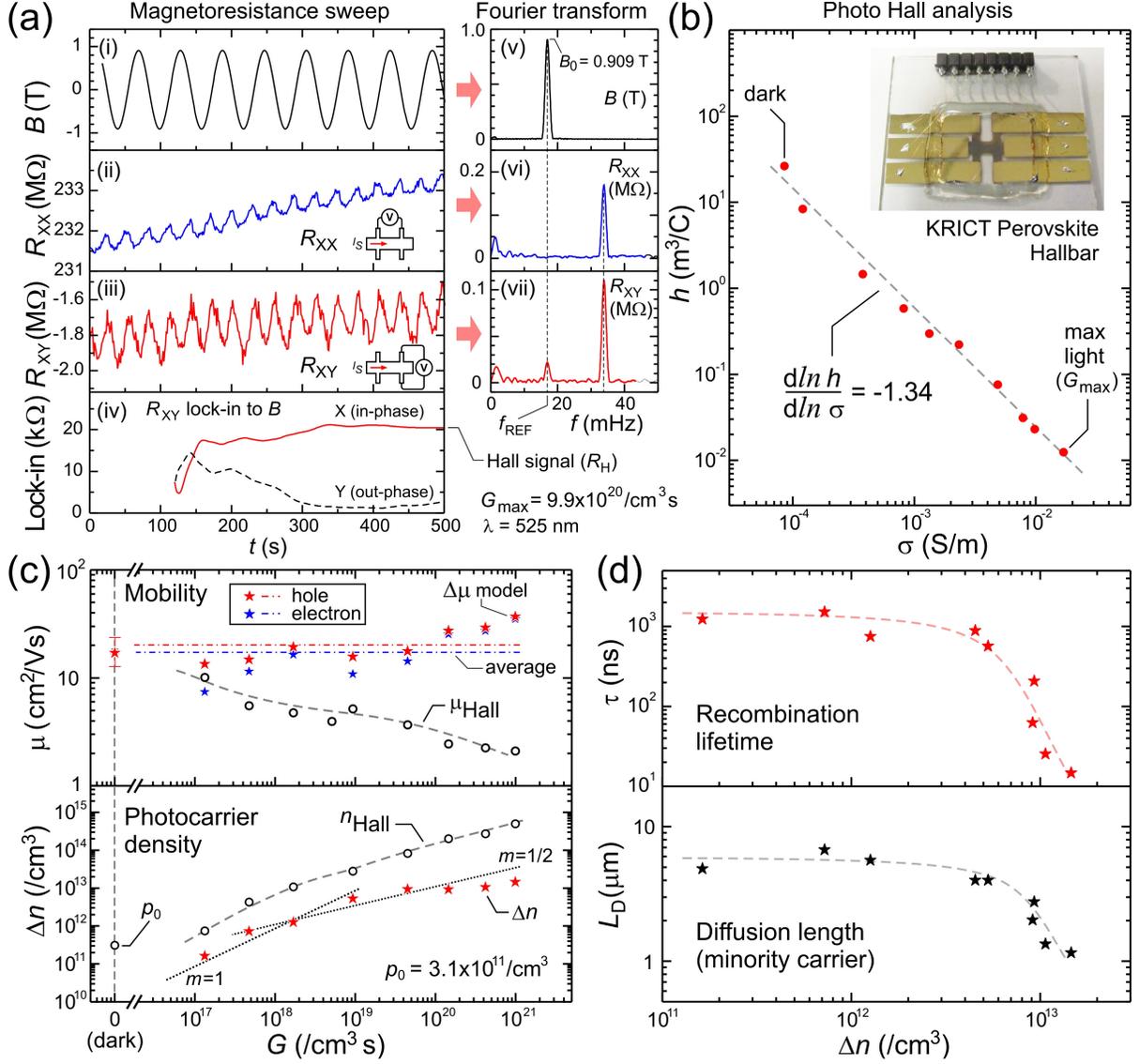

**Fig. 2 Carrier-resolved photo-Hall analysis in world-record-quality perovskite:** (a) Magneto-resistance sweep ($R_{XX}$: longitudinal and $R_{XY}$: transverse), Fourier transform and lock in detection of the Hall signal ($R_{XY}$). (b) $h$-$\sigma$ plot for photo-Hall analysis. Inset: the perovskite Hall bar device. (c) Results: Majority ($\mu_p$) and minority ($\mu_n$) mobility and photo-carrier density $\Delta n$ vs. absorbed photon density $G$. (d) Recombination lifetime and minority carrier diffusion length mapped against $\Delta n$ (all dashed curves are guides for the eye only).

We obtained a series of $\sigma$ and $h$ points that change significantly with illumination [Fig. 2(b)]: $\sigma$ increases by ~200× and $h$ drops by ~3000× (this perovskite study demonstrates the photo-Hall experiment in the high injection regime where $\Delta\sigma >> \sigma_0$, with $\sigma_0$ representing the dark conductivity). The new photo-Hall identity equation (Eq. 1) *provides an astonishingly simple and quick insight into the data* by looking at the slope of the $h$-$\sigma$ data using the log scale. If the slope is equal to -2 then $\mu_p = \mu_n$, if it is larger (less) than -2 then $\mu_p > \mu_n$ ($\mu_p < \mu_n$). Furthermore the $h$-$\sigma$ data can be plotted employing arbitrary units, e.g. using $\log(R_{XY})$ vs. $\log(1/R_{XX})$ to perform the same analysis. From Fig 2(b) we obtain the overall $d \ln h / d \ln \sigma = -1.34$, which implies that $\mu_p > \mu_n$. Furthermore we could evaluate $\Delta\mu$ at any $h$-$\sigma$ point, e.g. at the maximum light intensity: $\Delta\mu = 1.7$ cm$^2$/Vs. In practice we only need to know the local slope at the $h$-$\sigma$ point of interest; thus knowing only neighboring data points is sufficient (besides $p_0$ in the dark). We proceed to solve for $\mu_p$, $\mu_n$ and $\Delta n$ using this "$\Delta\mu$" model and plot them with respect to $G$ in Fig. 2(c). We use the sample thickness $d$ in calculating $\Delta n$, as $d < 1/\alpha + L_D$, where $\alpha$ is the absorption coefficient, which implies that the whole sample is populated by the photogenerated carriers. We obtain average mobility: $\mu_p = (20\pm8)$ cm$^2$/Vs, $\mu_n = (17\pm10)$ cm$^2$/Vs, and $\Delta n$ which increases with $G$ as expected. For comparison we also plot the "single carrier" Hall mobility (density): $\mu_{Hall} = h\sigma$ ( $n_{Hall} = 1/h\,e$ ), which is often used to estimate $\mu$ and $\Delta n$ in past photo-Hall studies (16). As seen in Fig. 2(c) the actual values $\mu_p$, $\mu_n$ and $\Delta n$ are very different from $\mu_{Hall}$ and $n_{Hall}$.

The $\Delta n$ vs. $G$ data in Fig. 2(c) allow us to investigate the recombination mechanism in the perovskite film with unprecedented detail. The data are consistent with two different regimes of recombination following a steady-state rate equation (16): $\eta G = \Delta n / \tau_{TR} + \gamma \Delta n^2$, where $\eta G$ is the generation rate, $\eta$ is the photocarrier generation efficieny, $\tau_{TR}$ is the trap limited carrier lifetime and $\gamma$ is the (bimolecular) recombination coefficient. These recombination mechanisms can be expressed as: $\Delta n \propto G^m$ where $m = 1$ (0.5) is associated with trap (bimolecular) recombination regime. Fit lines with $m = 1$ and 0.5 are also plotted in Fig. 2(c) and, from the the slope of the latter, we obtain $\gamma = (3.2\pm2)\times10^{-7}$ cm$^3$/s. Part of the uncertainty stems from the "effective thickness" $d_{eff}$ used to calculate $\Delta n$. The upper (lower) bound is associated with the sample thickness $d$ (absorption length) as $d_{eff}$. For theoretical comparison, the Langevin recombination model (23) predicts: $\gamma = e(\mu_p + \mu_n)/\varepsilon_r\varepsilon_0 = 3.8\times10^{-6}$ cm$^3$/s, where $\varepsilon_0$ and $\varepsilon_r = 18$ are the vacuum and relative (24) dielectric constant respectively.

Finally we can determine $\tau$ and $L_D$, using $\tau = \Delta n / \eta G$ and $L_D = \sqrt{k_B T \mu_n \tau / e}$, where $k_B$ is the Boltzmann constant, $T$ is temperature and we assume $\eta$ is close to unity. Furthermore, we can map these results as a function of $\Delta n$ as shown in Fig. 2(d). We obtain $\tau \sim (15\text{-}15{,}000)$ ns and $L_D \sim (1.2\text{-}6.7)$ μm; both show decreasing trends towards high light intensity as expected (25). We also compare our results with various recent transport studies in SM Table S3. In general, our results are consistent with these studies. However, we note that, given the wide variation of $\tau$ and $L_D$ with $G$ or $\Delta n$, indicating the illumination intensity when reporting lifetime measurements is crucial. The relatively long $\tau$ and $L_D$ obtained in this study attest to the high quality of this perovskite film (21).

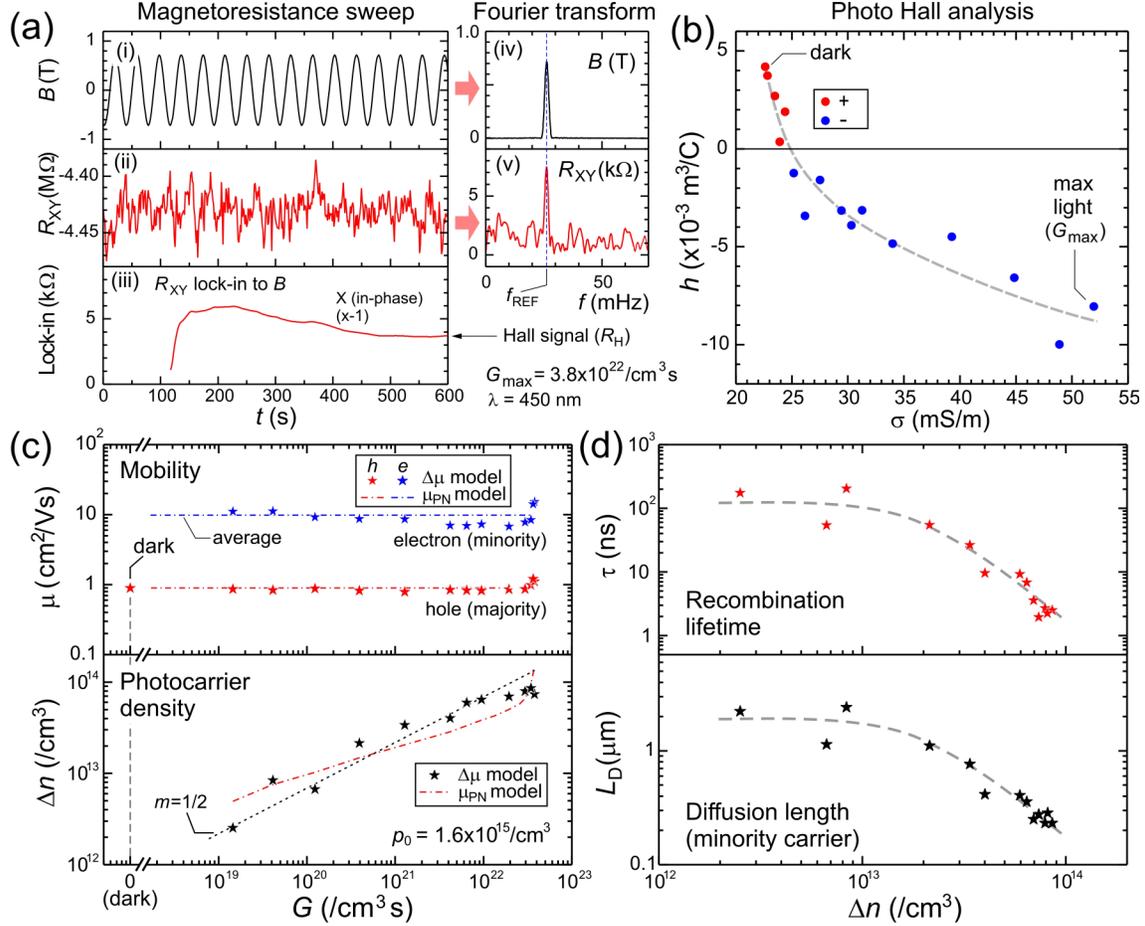

**Fig. 3 Carrier-resolved photo-Hall analysis in world-record-quality kesterite:** (a) Magneto-resistance sweep, Fourier transform and lock in detection of the Hall signal. (b) $h$-$\sigma$ plot for photo-Hall analysis showing inversion of $h$. (c) Results: Majority ($\mu_p$) and minority ($\mu_n$) mobility and photo-carrier density $\Delta n$ vs. absorbed photon density $G$. (d) Recombination lifetime and minority carrier diffusion length mapped against $\Delta n$ (all dashed curves are guides for the eye only).

We further show a study on a world-record-quality CZTSSe film (see SM A), which provides an interesting example of carrier-resolved photo-Hall analysis in the low injection regime ($\Delta\sigma \sim \sigma_0$), due to the much larger $p_0$ ($\sim 10^4\times$) compared to perovskite. Additionally, the sign of the Hall coefficient becomes inverted due to much higher minority (electron) mobility that dominates the conductivity at high light intensity. We perform the experiment on a CZTSSe van-der Pauw sample ($d$=2 μm) using a blue laser ($\lambda$= 450 nm, up to ~300 mW/cm²). In the dark, the sample shows $P$-type conductivity with $p_0$=(1.6±0.1)×10¹⁵ /cm³ and $\mu_p$=(0.9±0.1) cm²/Vs, whose low values compared to that of single crystal (*26*) may suggest grain-boundary-limited mobility. The $R_{XY}$ signal at high light intensity [Fig. 3(a)] is noisy, most likely as a result of dominant minority carrier conduction that strongly depends on light intensity (which is not stabilized). Nevertheless, the Fourier spectrum shows a clear and robust peak at $f_{REF}$ [Fig. 3(a)(v)], which produces a steady lock-in output $R_H$. This again highlights the importance of ac-Hall and lock-in detection to obtain clean Hall signals. We then plot the $h$-$\sigma$ curve in Fig. 3(b) and observe a monotonic behavior. As the change in conductivity is rather small, for analysis we plot the $h$-$\sigma$ curve on a linear scale and expand Eq. 1 to: $\Delta\mu = \sigma^2 dh/d\sigma + 2h\sigma$ (also we cannot evaluate $\ln h$ when $h$ is negative). The $h$ data are rather noisy due to low majority mobility, nevertheless we can construct a smoothly varying outline curve (dashed curve in Fig. 3(b)) from which we can determine the slope $dh/d\sigma$ and solve for $\Delta n$, $\mu_p$ and $\mu_n$ using the "$\Delta\mu$" model. We plot the results in Fig. 3(c) and obtain a significantly higher minority (electron) mobility: $\mu_n$=(9.8±2.6) cm$^2$/Vs compared to the majority: $\mu_p$=(0.9±0.1) cm$^2$/Vs. This "$\Delta\mu$" model yields a fairly constant $\mu_p$ but larger uncertainty in $\mu_n$ mainly due to uncertainty in $h$ and d$h$/d$\sigma$. For comparison, one could use the constant average mobility values obtained from the "$\Delta\mu$" model and solve for $\Delta n$ (the dash-dot red curve in Fig. 3(c)), which we refer to as the "$\mu_{PN}$" model. From $\Delta n$ vs. $G$ data we also obtain $\gamma$=(2.1-21)×10$^{-7}$ cm$^3$/s.

Similarly we calculate $\tau$ and $L_D$ as shown in Fig. 3(d) and note that the sample exhibits a very high mobility ratio (~11×), which explains the strong $h$ inversion at high light intensity. In a separate study on completed solar cell devices, we have also performed extraction of minority carrier mobility involving biased-QE to determine $L_D$'s and TR-PL to determine $\tau$'s (3). This study yielded: $\mu_n$~(10-260) cm$^2$/Vs, $\tau$~(1.5-16) ns, and $L_D$~(0.3-1.2) μm. Our photo-Hall study here yields consistent results with $\mu_n$~10 cm$^2$/Vs, $\tau$~(2-200) ns and $L_D$~(0.23-2.4) μm depending on $G$. We highlight that this study was done on a bare absorber layer, which is more accurate because of lack of charge carrier separation due to the junction, and more favorable for higher throughput characterization. In addition, this technique provides simultaneous extraction of all carrier parameters under the same light condition, which avoids the issues of different light illumination as encountered in our previous study. We compare our study with various recent transport studies in CZTSSe in SM Table S4.

Of the dozens of electrical transport measurements performed on perovskites and kesterites as summarized in Tables S3 and S4, this is the first time that all minority and majority carrier characteristics ($\mu$, $\Delta n$, $\tau$, $L_D$, $\gamma$) have been measured *simultaneously from a single experimental setup, on a single sample and mapped against varying light intensities*. This clearly demonstrates the power of this carrier-resolved photo-Hall technique, which significantly expands the capability of the Hall measurement since its original discovery in 1879 (27).

## Acknowledgements:

S.R.P. and B.S. thank financial support from the Technology Development Program to Solve Climate Changes of the National Research Foundation (NRF) funded by the Ministry of Science, ICT & Future Planning (No. 2016M1A2A2936757), and from the Global Frontier R&D Program on Center for Multiscale Energy System funded by NRF under the Ministry of Science, ICT & Future Planning, Korea (2012M3A6A7054855). J.H.N. thanks financial support from NRF grant funded by the Korea government (MSIP) (2017R1A2B2009676, 2017R1A4A1015022). D.B.M. thanks National Science Foundation for support under Grant No. DMR-1709294. We thank Supratik Guha for managing IBM PV program, Michael Pereira for hardware construction, and Jekyung Kim for Table S4. The PDL Hall system was developed at IBM Research and documented in the following patents: US 9,041,389 (*18*), US 9,772,385 (*19*) and application: US 15/281,968 (2016).

## Authors contributions:

O.G. (PI, IBM) and B.S. (PI, KAIST) conceived the project. O.G. built the experimental setup, programmed the analysis software, derived Eq. 1 and other formulas, performed measurement and analysis and led the manuscript writing. S.R.P. prepared samples, performed optical and Hall measurements and analysis. O.G, S.R.P., B.S., D.B, developed data analysis, interpretation and participated in manuscript writing. Y.V helped in PDL system development and formula derivation. Y.S.L and D.B. helped in optical study. N.J.J and J.H.N. prepared the perovskite samples and solar cells. D.B.M, T.T. and X.S. developed the champion CZTSSe process, D.B.M. managed IBM PV program and participated in manuscript writing.

## Supplementary Materials:

Supplementary Materials A-G.
Figures S1-S4
Tables S1-S4
Equations 4-25
Animation Movie S1
References (28-71)

# Carrier-Resolved Photo Hall Measurement in World-Record-Quality Perovskite and Kesterite Solar Absorbers

Oki Gunawan[1*], Seong Ryul Pae[2], Douglas M. Bishop[1], Yun Seog Lee[1,3], Yudistira Virgus[1], Nam Joong Jeon[4], Jun Hong Noh[4,5], Xiaoyan Shao[1], Teodor Todorov[1], David B. Mitzi[6], Byungha Shin[2*]

**This PDF file includes:**

> Supplementary Text
> Figs. S1 to S4
> Tables S1 to S4
> Caption for Movie S1

**Other Supplementary Materials:**

> Movie S1

# Contents:

# A. Materials and Methods

## (1) Perovskite film

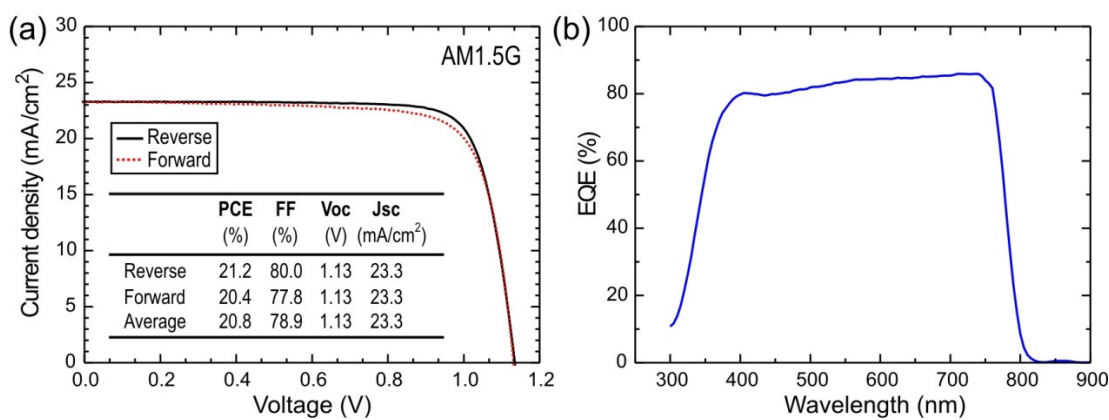

**Fig. S1** Device performance of the $(FAPbI_3)_{0.88}(MAPbBr_3)_{0.12}$ solar cell: (a) J-V curves measured by reverse and forward scans. The photovoltaic performance values are summarized in the inset. (b) The external quantum efficiency (EQE) spectrum.

The perovskite films for carrier-resolved photo-Hall measurement are based on the $(FAPbI_3)_{1-x}$ $(MAPbBr_3)_x$ mixed-perovskite system, and employ a halide perovskite composition analogous to that used in a previous world record power conversion efficiency (PCE) device reported by KRICT (*21*). After reporting a PCE of 17.9 % at x=0.15, the PCE has since been improved by modifying the film deposition method and adjusting the value of x. The detailed procedure for halide film formation and device fabrication is presented in next section. Using this approach, we demonstrated a high PCE ($\eta$) of 20.8 % for x=0.12 under 1 sun condition (AM1.5G, 100mW/cm$^2$), within a FTO/bl-TiO$_2$/mp-TiO$_2$/perovskite/PTAA/Au device structure. Figure S1(a) shows photocurrent density-voltage (*J-V*) curves for the $(FAPbI_3)_{0.88}(MAPbBr_3)_{0.12}$ device measured by reverse and forward scans with 10 mV voltage steps and 40 ms delay times under AM 1.5 G illumination. The device exhibits a short circuit current density ($J_{SC}$) of 23.3 mA/cm$^2$, open circuit voltage ($V_{OC}$) of 1.13 V, and fill factor (FF) of 80.0 % by reverse scan. A slight decrease in FF to 77.8% under forward scanning direction results in an average PCE of 20.8 %. The external quantum efficiency (EQE) spectrum for the device is presented in Figure S1(b), showing a very broad plateau of over 80 % between 400 and 750 nm.

**Film formation for photo-Hall measurement.** All precursor materials were prepared following a previous report (*21*). To form perovskite thin films based on the $(FAPbI_3)_{0.88}(MAPbBr_3)_{0.12}$ composition, the 1.05 M solution dissolving NH$_2$CH=NH$_2$I (FAI) and CH$_3$NH$_3$Br (MABr) with PbI$_2$ and PbBr$_2$ in *N-N*-dimethylformamide (DMF) and dimethylsulfoxide (DMSO) (6:1 v/v) was prepared by stirring at 60 $^o$C for 1 hour. Then, the solution was coated onto a fused silica substrate heated to 60 °C by two consecutive spin-coating steps, at 1000 and 5000 rpm for 5 and 10 s, respectively. During the second spin-coating step, 1 mL diethyl ether was poured onto the substrate after 5 s. Then, the substrate was heat-treated at 150 °C for 10 min. A compact $(FAPbI_3)_{0.88}$ $(MAPbBr_3)_{0.12}$ film with a thickness of 550 nm was obtained. Then, we selectively scraped the film off the substrate to pattern the desired Hall bar for photo-Hall measurement. The Hall bar has an active area of 2 mm × 4 mm as shown in the inset of Fig. 2(b).

**Device fabrication and characterization.** A 70 nm-thick blocking layer of $TiO_2$ (bl-$TiO_2$) was deposited onto an F-doped $SnO_2$ (FTO, Pilkington, TEC8) substrate by spray pyrolysis using a 10 vol% titanium diisopropoxide bis(acetylacetonate) solution in ethanol at 450 °C. A $TiO_2$ slurry was prepared by diluting $TiO_2$ pastes (ShareChem Co., SC-HT040) in mixed solvent (2-methoxyethanol:terpineol = 3.5:1 w/w). The 100 nm-thick mesoporous-$TiO_2$ (mp-$TiO_2$) was fabricated by spin coating the $TiO_2$ slurry onto the bl-$TiO_2$ layer and subsequently calcinating at 500 °C for 1 h in air to remove the organic components. Bis(trifluoromethane)sulfonimide lithium salt was treated onto the mp-$TiO_2$ layer. Then, the $(FAPbI_3)_{0.88}(MAPbBr_3)_{0.12}$ film was formed using the above-mentioned method. A polytriarylamine (PTAA) (EM index, $M$ n = 17 500 g mol$^{-1}$)/toluene (10 mg/1 mL) solution with an additive of 7.5 μL Li-bis(trifluoromethanesulfonyl) imide (Li-TFSI)/acetonitrile (170 mg/1 mL) and 4 μL 4- *tert* -butylpyridine (TBP) was spin-coated on the perovskite layer/mp-$TiO_2$ /bl-$TiO_2$ /FTO substrate at 3000 rpm for 30 s. The *J–V* curves were measured using a solar simulator (Newport, Oriel Class A, 91195A) with a source meter (Keithley 2420) at 100 mA cm$^{-2}$ AM 1.5G illumination and a calibrated Si-reference cell certificated by NREL. The *J–V* curves for the device were measured by masking the active area with a metal mask 0.094 cm$^2$ in area. The EQE was measured by a power source (Newport 300W Xenon lamp, 66920) with a monochromator (Newport Cornerstone 260) and a multimeter (Keithley2001).

## (2) Kesterite film

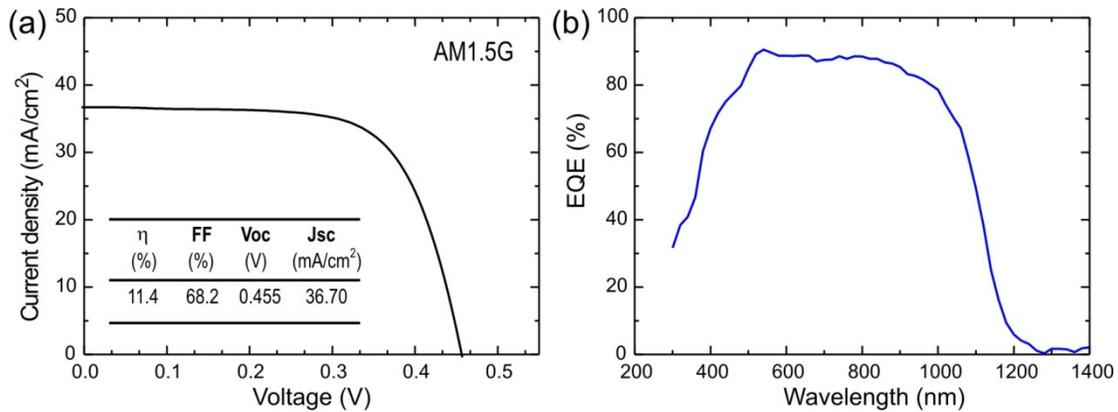

**Fig. S2** Device performance of the CZTSSe solar cell: (a) J-V curves under AM1.5G simulated light. The photovoltaic parameters are shown in the inset. (b) The EQE spectrum.

The CZTSSe film was prepared as previously reported using a hydrazine-based pure solution approach (*7*) The final thickness of the CZTSSe absorber layer was ~ 2μm. The CdS emitter was deposited by chemical bath deposition (CBD). Transparent conductive oxide (ZnO/ITO) was sputter deposited on the CdS. The electrical contact grid for the cell was made by evaporation (deposition area defined by a mask) of a thin layer of Ni followed by 2 μm of Al. A MgF anti-reflective coating was deposited by evaporation. The finished device was then scribed to isolate the cell and the area of the cell was measured using an optical microscope as $A = 0.453$ cm$^2$. The device characteristics of a high performing device from the same film batch as used for the photo-Hall study are shown in Fig. S2.

For photo-Hall study, the CZTSSe absorber layer was isolated from the blank area of a finished device after removing the top stack (CdS/ZnO/ITO/Ni/Al/MgF$_2$) by sonication in 10% aqueous HCl for 2 minutes and rinsing with water. Afterwards the absorber layer was exfoliated onto a secondary 5mm × 5mm glass substrate, from the underlying Mo/glass layer, using the method described in Ref. (*28*). Then we deposited four terminal Ti/Au (10nm/100nm) square contacts on the four corners to define a Van der Pauw sample with area of 3 mm × 3 mm.

## B. The Parallel Dipole Line Hall System

An ac magnetic field is critical for Hall measurements of materials with a low range of mobilities ($\mu < 10$ cm$^2$/Vs) or highly resistive samples like perovskites. Implementation of an ac field with traditional electromagnets is not practical due to the highly inductive impedance that attenuates the output field, unless a complex variable power factor correction with large capacitor banks is used. Commercial ac Hall electromagnet systems are available, but very expensive (> US$100k); thus, ac field generation systems using a permanent magnet have been pursued by many groups (*17-19,29,30*).

The parallel dipole line (PDL) Hall system was originally developed for CZTSSe development at IBM Research (*17*) and possesses natural characteristics suitable for high sensitivity ac-Hall /photo-Hall experiments. The PDL system is based on a recently-discovered natural magnetic trap that harbors a new type of field confinement effect from a camelback field profile along its central longitudinal axis (*17,20* ). The PDL Hall system consists of a pair of dipole line or diametric magnets, i.e. cylindrical magnets with uniform transverse magnetization. The diametric magnet produces an exterior magnetic field equal to that of a linear distribution of transverse dipole (*17*). This effect is analogous to the fact that a uniformly magnetized sphere produces a perfect point dipole field at its exterior.

To describe the field characteristics of a PDL system, we use the following elegant description suggested by K.T. McDonald (see Ref. (*20*), SM B). Consider a long dipole line (or a long diametric magnet) system, the field distribution can be described by a simple complex potential function of the form $f(z) \sim 1/z$, specifically:

$$f(z) = \frac{M}{2R^2} \frac{1}{z} \ ,$$ (4)

where $z = x + i\,y$ is a complex number within the $x$, $y$ coordinate system, $M$ is the volume magnetization of the magnet and $R$ is the magnet radius. The magnetic field outside the magnet is given as:

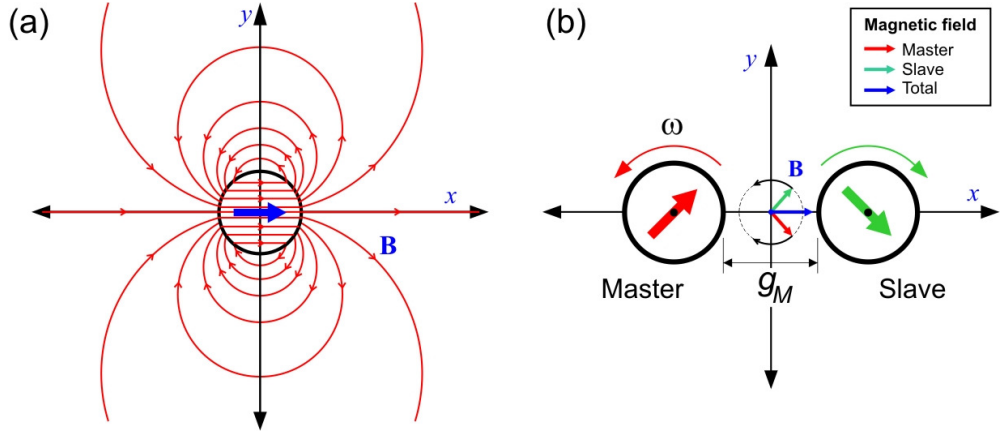

**Fig. S3** (a) A long dipole line (diametric) magnet and its field distribution. (b) The rotating PDL Hall system and its field evolution (see also Movie S1).

$$\mathbf{B}(x, y) = -\mu_0 \nabla \mathrm{Re}[f(z)], \tag{5}$$

which yields: $\mathbf{B}(x, y) = \mu_0 M R^2 / 2(x^2 + y^2)^2 \times [(x^2 - y^2)\hat{\mathbf{x}} + 2x y \hat{\mathbf{y}}]$, where $\mu_0$ is the magnetic permeability in vacuum. This field distribution is plotted in Fig. S3(a). Note that the field lines form parts of *perfect circles* (*20*).

In the rotating PDL Hall configuration, we have a pair of diametric magnets whose centers are placed at ($\pm a$, 0), separated by a gap $g_M$ as shown in Figure S3(b), where $a = R + g_M / 2$. The sample will be placed at the center, with the sample normal pointing along the $x$-axis. Only one of the magnets (the "Master") needs to be driven by a motor and gearbox system with angular speed $\omega$ and the other magnet (the "Slave") synchronously follows in the opposite direction. The complex potential function of such a rotating PDL system is given as:

$$f_T(z) = \frac{M}{2R^2}\left( \frac{e^{i\omega t}}{z - a} + \frac{e^{-i\omega t}}{z + a} \right) \tag{6}$$

The total magnetic field at the center (0, 0) where the sample resides reduces to a very simple form:

$$\mathbf{B}_T(0,0) = -\mu_0 \nabla \mathrm{Re}[f_T(z)] = B_0 \cos\omega t\, \hat{\mathbf{x}}, \qquad \text{with:} \ \ B_0 = \frac{\mu_0 M R^2}{(R + g_M / 2)^2} \tag{7}$$

Note that the $y$-component field vanishes, leaving only the $x$-component, and a pure single harmonic field term: $\cos\omega t$. The geometric description of the system also provides a simple explanation of this rotating PDL field characteristic as shown in Figure S3(b). Each magnet produces a counter-rotating field that cancels off the $y$-component, leaving behind only a total oscillating field in the $x$-direction. An animation of the PDL field evolution is shown in Movie S1, which provides clear visualization of this effect. *We emphasize that this natural field characteristic* of the rotating PDL system produces *the desired qualities for our photo-Hall experiment: a unidirectional field that mainly excites the desired Hall effect, without any out-of-plane field that may produce an extra parasitic magnetoresistance signal, and a pure harmonic ac field that simplifies the Fourier analysis and lock-in detection.* We also calibrate the magnetic field at the sample position (i.e., placed at the center of the PDL system) for various magnet gaps using a gaussmeter (Lakeshore 410 Gaussmeter). This then represents the amplitude ($B_M$) of the oscillating field.

Another important design consideration is the size or aspect ratio (length / radius) of the diametric magnet. If the length of the magnet is too short, the field on the sample will be very non-uniform, while if is it is too long the torque required to drive the system will be unnecessarily large (requiring a more powerful motor or gearbox and causing a more jerky motion due to the large oscillating torque). For this problem we have to recognize the *camelback effect*, which is the central feature of the PDL system that enables it to become a natural diamagnetic trap (*17*), as shown in Figure S4(a). This effect occurs when the length of the dipole line system exceeds a certain critical length $L_C$, which produces a slightly enhance field at the edges, thus producing a camelback field profile. Fig. S4(b) also shows the field distribution at the center of the PDL system with different magnet aspect ratios of $L/R = 1$, 2 and 4. The magnetic field along the $z$-axis is most uniform when the field profile at the center is flat ($d^2B/dz^2 = 0$). Therefore, we design our PDL magnet length *at the critical length for the camelback effect, which occurs at $L_C \sim 2.5a$, where $2a$ is the separation of the two dipole line system* (*20*). In the actual implementation, when we take into account a typical gap of ~5 mm, we choose $L \sim 2R$.

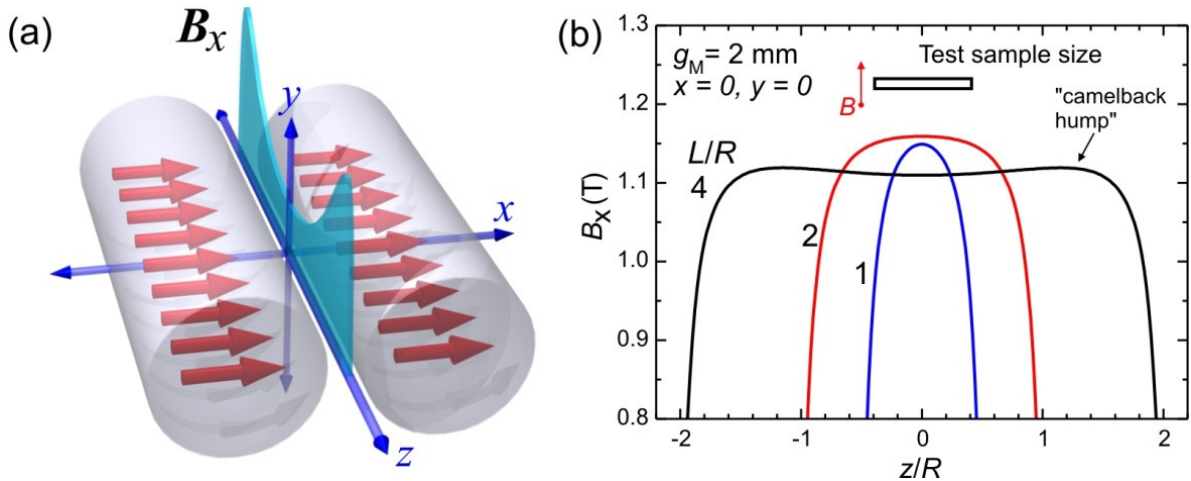

**Fig. S4** (a) The "camelback effect" (*20*) in a parallel dipole line system where the field ($B_x$) is enhanced at the edges. This effect is used to optimize the field uniformity in the PDL Hall system. (b) The field distribution along the longitudinal axis at various magnet aspect ratios: $L/R$ = 1, 2 and 4.

## C. The Photo-Hall Identity Equation and Related Formulas

Here we derive the photo-Hall identity equation starting from the well-known two-carrier Hall equations:

$$\sigma = e(p\,\mu_p + n\,\mu_n)\tag{8}$$

$$h = \frac{p - \beta^2 n}{(p + \beta n)^2 e}\tag{9}$$

where $\beta = \mu_n / \mu_p$ is the mobility ratio between the electron and hole. We use a *P*-type material as an example, but we emphasize that the photo-Hall identity equation also applies to *N*-type material. In a photo-Hall experiment in a *P*-type material we have: $p = p_0 + \Delta p$ and $n = \Delta n$. $\Delta n$ and $\Delta p$ are the electron and hole photo-carrier density respectively, which are equal under steady state equilibrium: $\Delta p = \Delta n$. As mentioned in the text, the key insight in solving the photo-Hall transport problem is to extract the information about the system mobilities from the $h$-$\sigma$ curve. The quantities $\sigma$ and $h$ are actually parametric as they are obtained experimentally as a function light intensity or absorbed photon density *G* or excess carrier density ($\Delta n$).

We have:

$$\sigma = e\,\mu_p\,[p_0 + \Delta n\,(1 + \beta)]\tag{10}$$

$$\frac{d\sigma}{d\Delta n} = e\,\mu_p\,(1 + \beta)\tag{11}$$

We assume that the dark carrier density $p_0$ and mobilities are fairly constant in the vicinity of the ($\sigma$,$h$) measurement point where the derivative is evaluated. We have the two-carrier Hall coefficient:

$$h = \frac{p_0 + \Delta n\,(1 - \beta^2)}{[p_0 + \Delta n\,(1 + \beta)]^2\,e} = \frac{e\,\mu_p^2\,[p_0 + \Delta n\,(1 - \beta^2)]}{\sigma^2}\tag{12}$$

$$\frac{d\,(h\,\sigma^2)}{d\Delta n} = e\,\mu_p^2\,(1 - \beta^2) = \frac{d\sigma}{d\Delta n}\,\frac{\mu_p\,(1 - \beta^2)}{1 + \beta}\tag{13}$$

$$\frac{dh}{d\sigma}\sigma^2 + 2\,h\,\sigma = \mu_p\,(1 - \beta) = \mu_p - \mu_n = \Delta\mu\tag{14}$$

Here is the key, the dependence on $\Delta n$ is eliminated; thus, we can relate the mobility difference only to $\sigma$ and $h$, which can be expressed more concisely as:

$$\Delta\mu = \left(2 + \frac{d\ln h}{d\ln\sigma}\right)h\,\sigma\ .\tag{15}$$

Repeating this derivation in an *N*-type material will lead to the same expression. Knowing $\Delta\mu$, we can then solve for the photo-Hall transport parameters ($\beta$, $\mu_p$, $\mu_n$ and $\Delta n$) by solving Eq. (9) and (8). These solutions are given in Table S1, Eq. (16) (17) (also shown in Eq. 2 and 3 in the main text) for *P*-type and *N*-type material respectively. We refer to this solution set as the "$\Delta\mu$" model. Note that one needs to know the dark or background carrier density, i.e. $p_0$ for *P*-type material or $n_0$ for *N*-type material, which is obtained from the classic Hall measurement in the dark.

In practice, at low light intensity when the sample resistance is very high or the conductivity is dominated by low mobility carriers, the Hall signal and thus *h* could be very noisy and thus the derivative term $d\ln h/d\ln\sigma$ may contain large uncertainties. In such situations, one can utilize the values that have been extracted using the "$\Delta\mu$" model at high light intensity or the overall average, and use another set formulas. For example one could use the known $\beta$, majority mobility ($\mu_M$) or both ("$\mu_{PN}$" for $\mu_p$ and $\mu_n$) and solve the transport problem accordingly. We refer to these as "$\beta$", "$\mu_M$" and "$\mu_{PN}$" models, respectively, as summarized in Table S1. For example, we use the average $\mu_P$ and $\mu_N$ ("$\mu_{PN}$" model) for comparison in the kesterite photo-Hall analysis in the text. These three models are useful for checking the overall solutions using a single average $\beta$ or set of average majority and minority mobility values. In all models, one need to know the dark or background carrier density or conductivity ($p_0$ / $n_0$, or $\sigma_0$).

**Table S1.** The carrier-resolved photo-Hall solution models and formulas for *P* and *N*-type materials

| NO | MODEL | TYPE | FORMULA | |
|----|-------|------|---------|---|
| 1. | $\Delta\mu$ | P | $\Delta\mu = \left(2 + \dfrac{d\ln h}{d\ln\sigma}\right) h\,\sigma$ | $\beta = \dfrac{2\sigma(\Delta\mu - h\,\sigma) - \Delta\mu^2 e\,p_0 \pm \Delta\mu\sqrt{e\,p_0}\sqrt{\Delta\mu^2 e\,p_0 + 4\sigma(h\,\sigma - \Delta\mu)}}{2\,\sigma(\Delta\mu - h\,\sigma)}$ <br><br> $\Delta n = \dfrac{\sigma(1-\beta) - e\,\Delta\mu\,p_0}{e\,\Delta\mu(1+\beta)}$ $\quad$ (16) |
| | | N | $\mu_p = \dfrac{\Delta\mu}{1-\beta}$ <br><br> $\mu_n = \beta\,\mu_p$ | $\beta = \dfrac{2\sigma(\Delta\mu - h\,\sigma) + \Delta\mu^2 e\,n_0 \pm \Delta\mu\sqrt{e\,n_0}\sqrt{\Delta\mu^2 e\,n_0 + 4\sigma(\Delta\mu - h\,\sigma)}}{2\,\sigma(\Delta\mu - h\,\sigma)}$ <br><br> $\Delta p = \dfrac{\sigma(1-\beta) - e\,\Delta\mu\,n_0\,\beta}{e\,\Delta\mu(1+\beta)}$ $\quad$ (17) |
| 2. | | P | $\Delta n = \dfrac{1 - 2\,e\,h\,p_0 - \beta \pm \sqrt{1 + \beta(\beta + 4\,e\,h\,p_0 - 2)}}{2\,e\,h(1+\beta)}$ <br><br> $\mu_p = \dfrac{\sigma}{e\,[\,p_0 + \Delta n(1+\beta)\,]}$ $\quad$ (18) | |

| | β | N | $$\Delta p = \frac{1-2ehn_0\beta(1+\beta)-\beta^2 \pm \sqrt{(1+\beta)^2(1+\beta[\beta-4ehn_0-2])}}{2eh(1+\beta)^2}$$ $$\mu_n = \frac{\sigma}{e[n_0 + \Delta p(1+\beta)/\beta]}$$ (19) |
|---|---|---|---|
| 3. | $\mu_M$ | P | $\beta = \frac{\sigma}{\mu_p}\frac{(\mu_p - h\sigma)}{(\sigma - ep_0\mu_p)}$ $\qquad\Delta n = \frac{\sigma - e\mu_p p_0}{e(1+\beta)\mu_p}$ (20) |
| | | N | $\beta = \frac{\mu_n}{\sigma}\frac{(\sigma - en_0\mu_n)}{(\mu_n + h\sigma)}$ $\qquad\Delta p = \frac{(\sigma - en_0\mu_n)\beta}{e\mu_n(1+\beta)}$ (21) |
| 4. | $\mu_{P-N}$ | P/N | $\Delta n = \frac{\sigma - \sigma_0}{e(\mu_p + \mu_n)}$ (22) |

## D. The Photo-Hall Measurement

The experimental setup is shown in Fig. 1(a). All measurements in this work were performed under ambient conditions. Photo-excitation was achieved by illumination using a solid state blue laser ($\lambda = 450$ nm, max. power 500 mW) or a Melles Griot He-Ne laser ($\lambda = 535$nm, max. power 10 mW). The sample is mounted at the center between the PDL magnets and a laser beam is directed to the sample through a motorized continuous neutral density filter, a cylindrical lens to expand the beam, a beam splitter and a wedge lens to deflect the beam onto the sample. The beam expander and wedge lens are used to obtain larger and more uniform illumination on the sample. A beam splitter was used to simultaneously illuminate the sample and a silicon "Monitor Photodetector" (PD) to monitor the photocurrent ($I_{PD-MON}$) at various light intensities. The $I_{PD-MON}$ is later used to determine the incident photon flux ($F$) or the absorbed photon density ($G$) on the sample, which will be discussed in the next section.

The electronics instruments consist of a custom-built PDL motor control box, custom-built decade shunt resistor (100k$\Omega$ to 10T$\Omega$) coupled with buffer amplifier for current measurement, Keithley 2400 Source Meter Unit (SMU) to apply the voltage or current source, Keithley 2001 Digital Multi Meter (DMM) for voltage measurement, Keithley 7065 Hall switch matrix card with high impedance buffer amplifiers for routing the signals between the samples, the SMU and DMM. For a high resistance sample like the perovskite we use the DC current excitation mode, and for lower resistance samples like kesterite we use the AC current excitation mode using a SRS830 lock-in amplifier. The PD current is measured using a Keithley 617 electrometer.

At every light intensity, we measure the sheet resistance ($R_S$) by performing the standard four-terminal Van der Pauw (six-terminal Hall bar) measurement by measuring 8 states (2 states) of longitudinal magnetoresistance (MR) $R_{XX}$. The conductivity of the sample is then calculated using: $\sigma = 1/R_S d$ where $d$ is the sample thickness. Next we perform the measurement for transverse MR: $R_{XY}$. The PDL master magnet is rotated by a stepper motor and gearbox system, typically with speed of 1 to 2 rpm, to generate the ac field on the sample. A Hall sensor is placed under the master magnet to monitor the oscillating field. The field oscillation and the $R_{XY}$ are then recorded as a function of time typically for 15 to 30 min each sweep. This measurement is repeated at several light intensities ranging from dark to the brightest condition using a motorized continuous neutral density filter, while recording the "Monitor PD" current to determine the incident photon flux density ($F$) or absorbed photon density ($G$) at each light condition. After all of the measurements are completed, the sample was replaced with a "Reference PD". We then determined the photo current ratio, $k_{PD} = I_{PD-MON} / I_{PD-REF}$, between the Reference PD and the Monitor PD at every given light intensity.

Once the measurement is done, we then performed the signal analysis using the PDL Hall analysis program, developed in MATLAB (*19*). We perform Fourier spectral analysis to inspect the existence of the MR signal ($R_{XY}$) at the same frequency as the magnetic field. We then proceed with phase sensitive lock-in detection, implemented by software, to extract the in-phase component of the Hall signal ($R_H$) while rejecting the out-of-phase component that arises from various sources such as Faraday emf induction. We use typical lock-in time constant of 120s - 300s. We then calculate the Hall coefficient given as, $h = R_{XY} d / B_M$, where $B_M$ is the magnetic field amplitude. Therefore at every light intensity, we obtain a set of $\sigma$ and $h$ values.

For the photo-Hall analysis, the calculated values of $\mu$, $\tau$ and $L_D$ are independent of the "effective thickness," $d_{eff}$, when the photo-carrier resides under steady state illumination as explained in the Supp. Mat. of Ref. (16). The effective thickness of this photo-carrier population is approximately $d_{eff} \sim 1/\alpha + L_D$ and it impacts the 3D photocarrier density calculation $\Delta n$. For a thin sample like our perovskite sample we have $d < 1/\alpha + L_D$, where $\alpha$ is the absorption coefficient, which implies that the photo generated carriers are occupying the whole sample and thus we can use the sample thickness $d$ as the effective thickness. In a thick sample where $d >> 1/\alpha + L_D$, one should use $d_{eff} \sim 1/\alpha + L_D$ in calculating $\Delta n$. This uncertainty in the effective thickness also impacts the calculated *e-h* recombination coefficient $\gamma$.

# E. Optical Property Determination

## (a) Transmittance and Reflectivity Measurement

Here we discuss the additional measurements, using the same setup shown in Fig. 1(a), to calculate the absorbed photon density, $G$ which requires the reflectivity, transmission and absorption coefficient. For transmittance and reflectivity measurements, the laser beam was deflected using a prism. For transmittance measurements, the sample under test deposited on glass substrates were mounted in-between a silicon photodetector ("PD Transmission") and the laser using an optical post. For reflectivity measurements, a silicon photodetector ("PD Reflectivity") embedded inside an integrating sphere (diameter ~80 mm) was used. The samples were placed on a wedge to partially deflect the beam into the integrating sphere where they are collected by the photodetector. The absorption coefficient of the films at the given wavelength λ can also be calculated using these reflectivity and transmission coefficients using the method detailed in Ref. (31).

### (b) Optical Constant Calculation

The absorbed photo density ($G$) in the sample is given by $G = F\,\alpha(\lambda)$ where $F$ is the photon flux density and $\alpha(\lambda)$ is the absorption coefficient at the operating wavelength. The impinging photon flux density, $F$, is related to the photocurrent reading from the monitor photodetector by:

$$F = \frac{I_{PD-MON}\,k_{PD}\,(1-R-T)}{e\,QE_{REF}(\lambda)\,A_{REF}} \quad, \tag{23}$$

where $QE_{REF}$ is the quantum efficiency of the Reference PD at wavelength $\lambda$, $A_{REF}$ is the effective area of the Reference PD (7.5 mm$^2$) and $k_{PD}$ is the PD calibration factor measured after the photo-Hall measurement session is completed. We define:

$$k_G = \frac{k_{PD}\,(1-R-T)\,\alpha(\lambda)}{e\,QE_{REF}(\lambda)\,A_{REF}}, \tag{24}$$

thus we can calculate $G$ directly from the Monitor PD current ($I_{PD\text{-}MON}$) in the experiment:

$$G = k_G\,I_{PD-MON}\,. \tag{25}$$

Below is the summary of the optical parameters for the perovskite and kesterite samples in this study:

**Table S2.** The optical parameters for the perovskite and kesterite sample in this study

| Sample | $\lambda$ (nm) | $k_{PD}$ | $QE_{REF}$ | $R$ | $T$ | $d$ ($\mu$m) | $\alpha$ (/m) | $k_G$ (/A m$^3$ s) |
|---|---|---|---|---|---|---|---|---|
| Perovskite | 525 | 1.11 | 0.633 | 0.244 | 0.0112 | 0.55 | $1.12 \times 10^7$ | $1.22 \times 10^{31}$ |
| Kesterite | 450 | 1.17 | 0.474 | 0.132 | 0.036 | 2 | $4.88 \times 10^6$ | $8.36 \times 10^{30}$ |

# F. Summary of Recent Perovskite Transport Studies

**Table S3.** Summary of recent electrical transport studies for lead(II) halide perovskites. The result from the current carrier-resolved photo-Hall effect study is highlighted in yellow.

| No | Refs | Material | Type | Techniques | Lifetime (ns) | Diffusion Length (μm) Electron | Hole | Mobility (cm²/Vs) Electron | Hole | Carrier conc. (/cm³) Electron | Hole | e-h RC γ (cm³/s) |
|---|---|---|---|---|---|---|---|---|---|---|---|---|
| 1 | (10) | $CH_3NH_3PbI_3$ | pc | PL, transient absorption. | 4.5±0.3 | 0.13 | 0.11 | | | | | |
| 2 | (11) | $CH_3NH_3PbI_{3-x}Cl_x$ | pc | Transient absorption, PL-quenching. | 273±7 | 1.07±0.20 | 1.21±0.24 | | | | | |
|  |  | $CH_3NH_3PbI_3$ | pc |  | 9.6 | 0.13±0.04 | 0.11±0.03 | | | | | |
| 3 | (32) | $CH_3NH_3PbI_3$ | pc | Intensity-modulated photocurrent/photovoltage. | | 1.2 - 1.5 | | | | | | |
| 4 | (8) | $CH_3NH_3PbI_3$ | pc | PL, transient absorption, time-resolved terahertz and microwave conductivity | | | | 12.5 | 7.5 | | | |
| 5 | (12) | $CH_3NH_3PbI_{3-x}Cl_x$ | pc | Ultrafast THz spectroscopy | | 2.7* | | 33* | | | | 1.1x10⁻¹⁰ |
| 6 | (9) | $CH_3NH_3PbI_{3-x}Cl_x$ | pc | Conductivity measurement. Total mobility used (e + h), assuming similar mass | | | | 20 | 20 | | | |
| 7 | (33) | $CH_3NH_3PbI_3$ | pc | Laser-flash time-resolved microwave conductivity | | | | 3 | 17 | | | 8x10⁻¹² |
| 8 | (34) | $CH_3NH_3PbI_3$ | pc | Hall effect | | | | 13.7-36.0 | | (2.4-5.9)x10¹⁴ | | |
| 9 | (29) | $CH_3NH_3PbI_3$ | pc | Hall effect | | | | 3.9 | | 2.8x10¹⁷ | | |
| 10 | (13) | $CH_3NH_3PbI_3$ | pc | Transient PL | 9 | | | | | | | |
|  |  | $CH_3NH_3PbI_{3-x}Cl_x$ | pc |  | 80 | | | | | | | |
| 11 | (35) | $CH_3NH_3PbI_3$ | pc | TR-PL, transient absorption | 140 | | | | | | | 1.7x10⁻¹⁰ |
| 12 | (36) | $CH_3NH_3PbI_3$ | pc | Transient THz spectroscopy | | ~1* | | 8.1* | | | | 9.2x10⁻¹⁰ |
|  |  | $CH_3NH_3PbI_{3-x}Cl_x$ | pc |  | | ~1* | | 11.6* | | | | 8.7x10⁻¹¹ |
| 13 | (14) | $CH_3NH_3PbI_3$ | sc | PL, transient absorption, Hall effect, time-of-flight measurement, space-charge-limited current technique | 22-1032 | 8* | | 2.5* | | 2x10¹⁰ | | |
|  |  | $CH_3NH_3PbBr_3$ | sc |  | 41-357 | 17* | | | 20-115 | 5x10⁹⁻¹⁰ | | |
| 14 | (15) | $CH_3NH_3PbI_3$ | pc | Transient photovoltai and impedance spectroscopy, Hall effect | 82-95 | | 175±25 | 24±4.1 | 164±25 | (9±2)x10⁹ | | |
| 15 | (37) | $CH_3NH_3PbI_{3-x}Cl_x$ | pc | Confocal fluorescence microscopy, PL | 1005 | | | | | | | 7.8x10⁻¹¹ |
| 16 | (38) | $CH_3NH_3PbI_3(Cl)$ | pc | Electron-beam-induced current | 50-100 | 0.36±0.02 | | | | | | |
| 17 | (39) | $FAPbBr_3$ | pc | THz transient photoconductivity | | 1.3* | | 14±2* | | | | ~1.0x10⁻⁹ |
|  |  | $FAPbI_3$ | pc |  | | 3.1* | | 27±2* | | | | ~1.0x10⁻¹⁰ |
| 18 | (40) | $CH_3NH_3PbI_3$ | pc | Time resolved microwave conductivity | | | | 29±6* | | | | |
| 19 | (41) | $FAPbI_3$ | sc | Space-charge-limited current | | 6.6* | | | 35 | 3.9x10⁹ | | |
|  |  | $FAPbBr_3$ | sc |  | | 19* | | | 62 | 1.5x10⁹ | | |
| 20 | (16) | $CH_3NH_3PbI_3$ | pc | Photo-Hall, photoconductivity. | 3x10⁴ | 23* | | 8* | | 9.0x10¹⁴ | | (1-5)x10⁻¹¹ |
|  |  | $CH_3NH_3PbBr3$ | pc |  | 3x10⁶ | 650* | | 60±5* | | 4.6x10¹² | | 8x10⁻¹¹ |
| 21 | (42) | $CH_3NH_3PbI_3$ | sc | TR-PL, time-resolved microwave conductance | 15,000 | 8 | 50 | | | | | 5.5x10⁻⁹ |
| 22 | (25) | $CH_3NH_3PbI_3$ | pc | Transient photovoltage, charge carrier extraction under 1 sun. | 390 | | | | | | 9.4x10¹⁸ | |
|  |  | $CH_3NH_3PbX_3$ | pc | (Carrier density under illumination) | 560 -1100 | | | | | | (2.4-5.9)x10¹⁹ | |
| 23 | (43) | $CH_3NH_3PbI_3$ | pc | Hall and magnetoresistance. (Mobility values under illumination) | 29,000 | 230* | | 68* | | 9.3x10¹¹ | | |
|  |  | $CH_3NH_3PbBr_3$ | pc |  | 36,000 | 24* | | 6.3* | | 4.3x10¹² | | |
| 24 | (44) | $FAPbI_{3-x}Cl_x$ | pc | Flash-photolysis time-resolved microwave conductivity. **Sum of mobility | 2,800 | | | 71** | | | | |
| 25 | (FAPbI3)1-x (MAPbBr3)x | | pc | Carrier-resolved photo-Hall effect. Values are mapped against G or Δn (Fig. 2) (IBM/KAIST/KRICT) | 15 - 1500 τ (G) | 1.2 - 6.7 $L_{D,N}(G)$ | 1.2 - 7.6 $L_{D,P}(G)$ | 17±9 $\mu_n(G)$ | 20±8 $\mu_p(G)$ | 3.1x10¹¹ Δn(G) | $p_0+\Delta n(G)$ | (1-5)x10⁻⁷ |

Notes: (1) * Carrier type is not specified (not-resolved). (2) pc = poly crystalline, sc = single crystal
(3) e-h RC γ = Electron-hole (bimolecular) recombination coefficient γ

In contrast to the classic Hall effect that only yields three information (majority carrier type, density and mobility), the carrier-resolved Photo Hall technique yields $6N+1$ values where *six* are associated with $\mu_p$, $\mu_n$, $\Delta n$, $\tau$, $L_{D,N}$ and $L_{D,P}$ repeated at $N$ light intensity settings and an extra parameter: the *e-h* recombination coefficient $\gamma$. In the main text, only the minority carrier diffusion length is reported ($L_D$).

# G. Summary of Recent Kesterite Transport Studies

**Table S4.** Summary of recent electrical transport studies for kesterites. The result from the current carrier-resolved photo-Hall effect study is highlighted in yellow. A study on the 12.6% PCE champion CZTSSe is reported in Row#14 (in bold).

| No | Refs | Material | Type | Techniques | Lifetime (ns) | Diffusion Length (μm) Electron | Hole | Mobility (cm²/Vs) Electron | Hole | Carrier conc. (/cm³) Electron | Hole | *e-h* RC γ (cm³/s) |
|---|---|---|---|---|---|---|---|---|---|---|---|---|
| 1 | (45) | CZTSe | pc | TR-PL | 7-9 | | | | | | | |
| 2 | (46) | CZTSSe | pc | TR-PL, Drive-level capacitance profiling (DLCP) | 3.1 | | | | | | $8 \times 10^{15}$ | |
| 3 | (47) | CZTSSe | pc | TR-PL | 10 | | | | | | | |
| | | CZTSe | pc | TR-PL | 12 | | | | | | | |
| 4 | (48) | CZTSSe | pc | DLCP | | | | | | | $(0.2\text{-}1.5)\times 10^{16}$ | |
| 5 | (49) | CZTSe | sc | Hall effect | | | | | 40-55 | | $1 \times 10^{17}$ | |
| 6 | (50) | CZTSe | pc | TR-PL | 6.7±1.4 | | | | | | | |
| 7 | (3) | CZTSSe | pc | TR-PL, EQE, reflectance measurement, capacitance-voltage (C-V) measurement | 1.5-15 | 0.3-1.2 | | 11-257 | | | | |
| 8 | (51) | CZTSSe | pc | TR-PL | 5 | | | | | | | |
| 9 | (52) | CZTS | pc | TR-PL, DLCP | 7.8 | ~0.35 | | | | | $1 \times 10^{16}$ | |
| 10 | (53) | CZTS | pc | TR-PL | 18.4 | | | | | | | |
| | | CZTSe | pc | | 9.3 | | | | | | | |
| 11 | (54) | CZTSe | pc | Hall effect | | | | | 1.23 | | $2 \times 10^{17}$ | |
| 12 | (55) | CZTSSe | pc | C-V | | | | | | | $(1\text{-}4)\times 10^{16}$ | |
| 13 | | CZTS | pc | Hall effect | | | | | | | $(4.9\text{-}6.9)\times 10^{15}$ | |
| 14 | **(7)** | **CZTSSe** | **pc** | **C-V, DLCP, EQE (PCE 12.6% champion)** | | **0.75±0.15** | | | | | **$8 \times 10^{15}$** | |
| 15 | (57) | CZTSe | pc | DLCP, EQE, reflectance measurement, TR-PL | 2-2.5 | 2.1±0.5 | | 690 | | | $2 \times 10^{15}$ | |
| 16 | (17) | CZTSSe | pc | Hall effect (ac/PDL) | | | | | 0.3-0.4 | | $(1.6\text{-}2.2)\times 10^{16}$ | |
| 17 | (58) | CZTSe | pc | TR-PL, DLCP | 4.2 | | | | | | $\sim 10^{16}$ | |
| 18 | (59) | CZTSe | pc | Hall effect | | | | | 0.2-0.5 | | $(2\text{-}20)\times 10^{16}$ | |
| 19 | (26) | CZTSe | sc | Hall effect | | | | | 50-150 | | $10^{17}\text{-}10^{19}$ | |
| 20 | (60) | CZTSSe | pc | TR-PL | 8.1 | | | | | | | |
| 21 | (61) | CZTSSe | pc | TR-PL | 2.6 | | | | | | | |
| | | CZTGeSSe | pc | | 10 | | | | | | | |
| | | AgCZTSe | pc | | | | | | | | | |
| 22 | (62) | CZTSe | pc | TR-PL | 2.4 | | | | | | | |
| | | AgCZTSe | pc | | 3.6-10 | | | | | | | |
| 23 | (63) | CZTSe | pc | Hall effect (ac/PDL), DLCP | | | | | 0.2 | | $1 \times 10^{16}$ | |
| | | AgCZTSe | pc | | | | | | 0.2-1.1 | | $10^{13}\text{-}10^{16}$ | |
| 24 | (64) | CZTSe | pc | TR-PL | 1.3-2.2 | | | | | | | |
| 25 | (65) | CZTS | pc | TR-PL,C-V | 5.8-12.4 | | | | | | $(3\text{-}5)\times 10^{16}$ | |
| 26 | (66) | CZTSSe | pc | C-V | | | | | | | $(1\text{-}2)\times 10^{16}$ | |
| 27 | (67) | CZTS | pc | TR-PL, DLCP | 4.1 | | | | | | $4.4 \times 10^{16}$ | |
| | | CZCTS | pc | | 10.8 | | | | | | $1.4 \times 10^{16}$ | |
| 28 | (68) | CZTS | pc | TR-PL, C-V | 10 | ~0.35 | | | | | $3 \times 10^{16}$ | |

| | | | | | $\tau$ | $L_{D,N}$ | $L_{D,P}$ | $\mu_n$ | $\mu_p$ | $\Delta n$ | $p_0+\Delta n$ | |
|---|---|---|---|---|---|---|---|---|---|---|---|---|
| 29 | (69) | CZTSe | pc | TR-PL,C-V | 1.6 | | | | | | $1.5\times10^{16}$ | |
| 30 | (70) | CZTSSe | pc | Hall effect (ac/PDL) | | | | | 0.3 | | $1\times10^{15}$ | |
| 31 | (71) | CZTSe | pc | TR-PL,C-V | 1-8 | | | | | | $10^{15}\text{-}10^{16}$ | |
| **32** | | **CZTSSe** | **pc** | **Carrier-resolved photo-Hall effect** Values are mapped against $G$ or $\Delta n$ (Fig. 3) **(IBM/KAIST/KRICT)** | **2-200** $\tau\,(G)$ | **0.23-2.4** $L_{D,N}(G)$ | **0.08-0.66** $L_{D,P}(G)$ | **9.8±2.6** $\mu_n(G)$ | **0.9±0.1** $\mu_p(G)$ | $\Delta n(G)$ | **1.6x10^{15}** $p_0+\Delta n(G)$ | **(2-21)x10^{-6}** |

Notes:     (1) pc = poly crystalline, sc = single crystal

(2) e-h RC $\gamma$ = Electron-hole (bimolecular) recombination coefficient $\gamma$

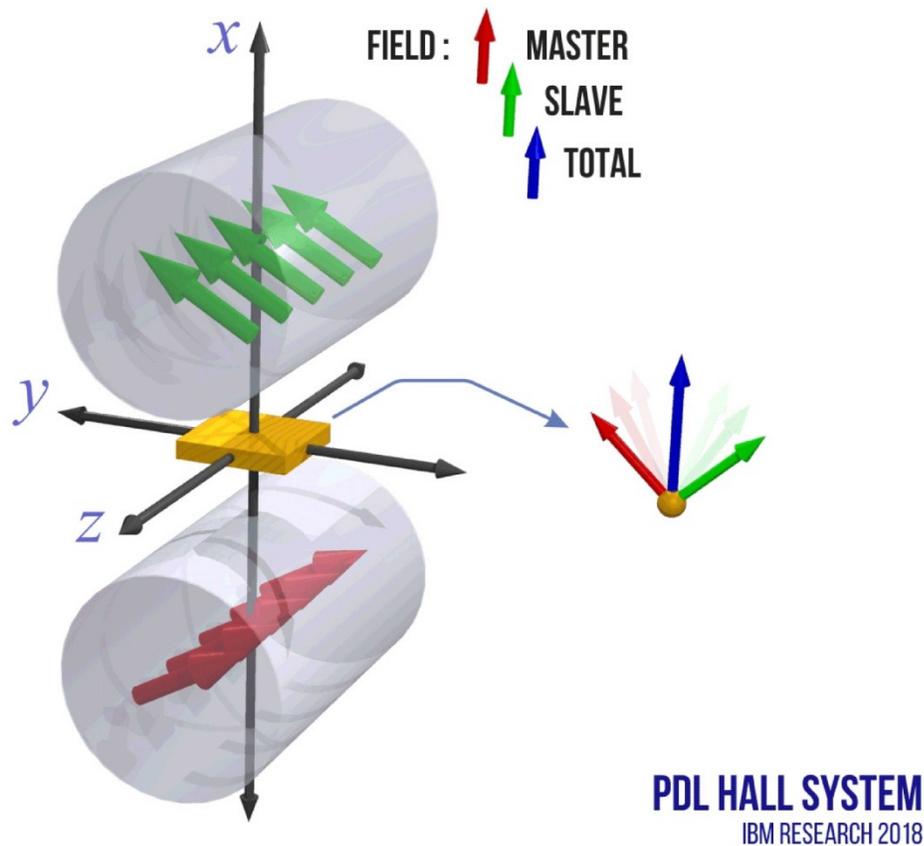

**Movie S1. Animation of the rotating parallel dipole line (PDL) Hall system and its field evolution.** The master magnet generates a counterclockwise field rotation (red) while the slave magnet follows synchronously but in opposite direction, generating a clockwise field rotation (green). This results in a total field (blue) which is *unidirectional* (always pointing normal to the sample) and *single harmonic* at the center where the sample resides.

**Movie link:** https://researcher.watson.ibm.com/researcher/files/us-ogunawa/Movie_S1_PDLHall.gif